\documentclass[useAMS,usenatbib]{mnras}
\usepackage{times}
\usepackage{natbib}
\usepackage{amssymb,amsmath}
\usepackage{graphicx,color}
\usepackage{hyperref}
\usepackage[usenames,dvipsnames]{xcolor}
\usepackage[OT2,T1]{fontenc}
\usepackage[draft]{todonotes}
\usepackage{multirow}
\usepackage{mathtools}
\usepackage[normalem]{ulem}
\usepackage{lineno}
\usepackage{gensymb}
\usepackage{setspace}
\usepackage{tabularx}
\usepackage{soul}
\usepackage{bm}
\usepackage{ulem}
\title[Weak Lensing Polarization]{The  Impact  of  Light  Polarization  Effects  on  Weak  Lensing  Systematics}

\author[C.-H. Lin et al.]
{\parbox{\textwidth}{Chien-Hao Lin$^1$\thanks{E-mail: \texttt{chienhal@andrew.cmu.edu}},
Brent Tan$^2$,
Rachel Mandelbaum$^1$,
Christopher M. Hirata$^3$
}
\vspace{0.4cm}\
\\$^1$McWilliams Center for Cosmology, Department of Physics, 
Carnegie Mellon University, Pittsburgh, PA 15213, USA,
\\$^2$Department of Physics, University of California
Santa Barbara, Santa Barbara, CA 93106, USA,
\\$^3$Center for Cosmology and Astroparticle Physics, The Ohio State University, Columbus, OH 43210, USA
}
\begin{document}
\date{\today}
\maketitle

\begin{abstract}
A fraction of the light observed from edge-on disk galaxies is polarized due to two physical effects: selective extinction by dust grains aligned with the magnetic field, and scattering of the anisotropic starlight field. Since the reflection and transmission coefficients of the reflecting and refracting surfaces in an optical system depend on the polarization of incoming rays, this optical polarization produces both (a) a selection bias in favor of galaxies with specific orientations and (b) a polarization-dependent PSF. In this work we build toy models to obtain for the first time an estimate for the impact of polarization on PSF shapes and the impact of the selection bias due to the polarization effect on the measurement of the ellipticity used in shear measurements. In particular, we are interested in determining if this effect will be significant for WFIRST. We show that the systematic uncertainties in the  ellipticity components are $8\times 10^{-5}$ and $1.1 \times 10^{-4}$ due to the selection bias and PSF errors respectively. Compared to the overall requirements on knowledge of the WFIRST PSF ellipticity ($4.7\times 10^{-4}$ per component), both of these systematic uncertainties are sufficiently close to the WFIRST tolerance level that more detailed studies of the polarization effects or more stringent requirements on  polarization-sensitive instrumentation for WFIRST are required.
\end{abstract}


\section{Introduction}
\label{sect:intro}
Weak gravitational lensing arises due to deflection of light by the gravitational fields of large-scale structure, leading to tangential shear distortions in galaxy shapes. Measuring the correlation functions of galaxy shapes is therefore a method by which we can measure the growth 
of structure in the Universe \citep[][]{2001PhR...340..291B,2007ApJS..172..239M, 2008ARNPS..58...99H, 2015RPPh...78h6901K, doi:10.1146/annurev-astro-081817-051928} and hence a powerful method for constraining cosmological parameters \citep[][]{ heh14,2016ApJ...824...77J, 2018arXiv181206076H, 2018PhRvD..98d3528T,2019PASJ..tmp...22H}.  

Since weak gravitational shear is only a percent-level signal, weak lensing measurements rely on the use of large galaxy samples to reduce statistical uncertainties. In the upcoming Stage-IV surveys, including Euclid\footnote{\url{http://sci.esa.int/euclid/}} \citep{2011arXiv1110.3193L}, LSST\footnote{\url{http://www.lsst.org/lsst}} \citep{2008arXiv0805.2366I, 2009arXiv0912.0201L}, and WFIRST\footnote{\url{https://wfirst.gsfc.nasa.gov}} \citep{2015arXiv150303757S}, the statistical uncertainties of the weak lensing measurements are expected to reach sub-percent level precision. With such small statistical uncertainties, the future of weak lensing analysis requires a better understanding and more careful control of systematics to avoid systematic uncertainties dominating over statistical uncertainties. Low-level sources of systematic uncertainty that are presently ignored in existing weak lensing analyses will become significant as the precision of the measurements increases. 

 Optical gravitational lensing measurements are based on how the light intensity profile of galaxies is affected by the matter distribution in the Universe. Besides intensity, the observed radiation also includes polarization information \citep{1989QJRAS..30..181R}. However, to the best of our knowledge, the information in polarization has not been considered in optical weak lensing. 
 While the light intensity profile contains the weak lensing information that we hope to measure, it is possible that the polarization of light could introduce weak lensing systematic errors if not accounted for.

Many polarization measurements of light from galaxies have been made at radio frequencies \citep{2002A&A...391...83B, 2007EAS....23...19B, Stil_2009, 2018PASJ...70R...2A}. 
At the radio frequency, the radiation is dominated by synchrotron emissions. Synchrotron emission is intrinsically linearly polarized perpendicular to the magnetic field; at long wavelengths, Faraday rotation changes the polarization angle and provides an additional probe of the magnetic field structure.

Weak lensing surveys are conducted, however, in the optical and near-infrared (NIR) wavelengths. The physics behind radio polarization is different from the optical/NIR. 
Unlike the radio band which is dominated by synchrotron emissions, the light at the optical and near infrared frequencies is not itself linearly polarized.
Some polarization effects may be induced through the interactions with dust grains \citep{1990ARA&A..28...37M}. The optical galactic polarization has been studied and measured in \cite{1990IAUS..140..245S,1991MNRAS.252P..12S,1996A&A...308..713F,1987MNRAS.224..299S,1998A&A...335..123F,1996QJRAS..37..297S,2000AJ....120.2920J}.
Even though the linear polarization of light from galaxies has been observed, polarization-induced systematic errors have not generally been considered for optical weak lensing analyses.

The overall weak lensing analysis covers many steps from image processing to the inference of cosmological parameter constraints. Different systematic uncertainties enter in each step. The sheared galaxy image that contains weak lensing information first propagates through the atmosphere (for ground-based observations) and the telescope optics, and is affected by the Point Spread Function (PSF) of both. The final images 
are further affected by detector non-idealities. The subsequent processes in the weak lensing analysis involve using the final image to construct the shear catalogues and summary statistics such as two-point correlation functions, and finally inferring the cosmological parameters using a likelihood analysis. Among all of these steps, the polarization effect influences how the light rays propagate in the optical system before the final images are realized on the detectors. Since the s and p-polarized components of light have different reflection and refraction coefficients at optical surfaces, including the mirrors and the anti-reflection coating on the detectors, the throughput of the optics and telescope PSF depend on the polarization of the input light \citep{2015PASP..127..445B,RC_polarization2015}. If the light from galaxies is linearly polarized, the existing analysis processes that ignore the polarization information could lead to biases in the weak lensing results. In this paper, we consider two potential polarization-related systematic errors: galaxy selection biases, and the PSF modeling errors due to the polarization-dependent PSF. The selection bias arises from the dependence of the transmitted intensity of polarized light on the angle of polarization. The polarization-dependent PSF, on the other hand, is a result of the polarization-dependent optical aberrations. 

In Section~\ref{sect:polarization}, we briefly review polarization effects in optical/NIR galaxy images. Sections~\ref{sect:selection_bias} and~\ref{sect:psf} describe the assumptions we make and how we construct our toy models for polarization-related selection and PSF effects respectively. We present our results in Section~\ref{sect:results} and conclude in Section~\ref{sect:conclusions}.


\section{Optical/NIR Polarization of galaxies}
\label{sect:polarization}

As shown in previous studies on optical polarization, light from distant stars and galaxies can be linearly polarized in the optical wavelengths \citep{1996A&A...308..713F, 1987MNRAS.224..299S, 1998A&A...335..123F}. At optical and near infrared frequencies, the linear polarization of light arises from physical mechanisms including the anisotropic scattering by spherical dust grains and the selective extinction by aligned dust grains. The alignment was once proposed to be caused by magnetic dissipation -- the Davis-Greenstein mechanism \citep{1951ApJ...114..206D} -- but is now attributed to precession around the magnetic field combined with torques from scattering of starlight, which also usually results in alignment of the long axis of the grains perpendicular to the magnetic field \citep{2015ARA&A..53..501A}. Both of these mechanisms can have complicated wavelength dependence. The dust scattering cross section increases toward the blue and peaks in the ultraviolet \citep{2003ApJ...598.1017D}, but the polarization fraction of the scattered light exhibits a non-monotonic behaviour \citep{2001ApJ...548..296W}. Polarization by selective extinction by aligned dust grains has long been known to peak in $\sim V$ band \citep[e.g.][]{1978ppim.book.....S}. By observing in four bands across the 0.9--2.0 $\mu$m observer-frame wavelength range, and studying sources across a range of redshifts, WFIRST will be sensitive to both sources of polarization in multiple regimes. 

\subsection{Toy Model for Galaxy Polarizations}

To simplify the calculation and make an order of magnitude estimation of the polarization effects, we begin by assuming that only edge-on disk galaxies are observed and that each edge-on galaxy has fixed polarized fraction $p$, which specifies the fraction of light from galaxy that is polarized. Although these assumptions will be relaxed by correction factors discussed in Sect.~\ref{sec:additional_factors}, this simplified model provides an easy and intuitive way to describe the galaxy polarization.

In the case of face-on disc galaxies, the polarization orientation varies with location in the galaxy, leading to the cancellation of optical polarizations in spiral and circular patterns \citep{1990IAUS..140..245S, 2000MNRAS.319..497S}. 
Thus, for a simple order of magnitude estimate of the polarization-dependent selection bias and PSF, we consider only edge-on disc galaxies in this work.
\cite{1996A&A...308..713F} reported the polarization of 3 edge-on galaxies, the optical polarization orientations of which are perpendicular to the major axes. However, it is also possible that the polarization orientation of an edge-on galaxy is parallel to the disk plane, as shown in \cite{1990IAUS..140..245S} and \cite{1996QJRAS..37..297S}.
The diversity of polarization orientation arises from two competing effects: the selective extinction of non-spherical dust grains aligned with the galactic magnetic field and the anisotropic scattering. The toroidal magnetic field on the disc plane would generate polarizations parallel to the major axis for edge-on galaxies (since the grains align perpendicular to the disc and preferentially absorb that polarization), while the polarization by anisotropic scattering is perpendicular to the major axes (since the grains are in a radiation field where more light is coming from directions parallel to than perpendicular to the disc). In \cite{Jones_2012}, the authors presented the integrated polarization survey of 70 nearby galaxies and suggested that the dust scattering is the dominant source of optical polarization.  
The magnitude of the polarization-dependence of the PSF is only affected by the polarization level rather than the polarization orientation, but the polarization orientation has a great impact on the selection bias. In our toy model, we consider only the polarization orientation that is perpendicular to the major axis of the edge-on galaxies. This will provide the upper limit for the polarization-induced selection bias.

The level of polarization is typically of order of several percent \citep{1986ARA&A..24..459S,1990IAUS..140..245S, 1995MNRAS.277.1430D,1996QJRAS..37..297S,Jones_2012}. 
For edge-on galaxies, the anisotropic scattering \citep{1982ApJ...258...59J} and alignment with the magnetic field \citep{1996A&A...308..713F, 1990IAUS..140..245S} under several $\mu$G of interstellar magnetic field both demonstrate polarization of order of 5\%. The integrated polarization level would be less than the level in polarization maps in \cite{1996A&A...308..713F} and \cite{ 1996QJRAS..37..297S} due to partial cancellation of polarization vectors over the galaxy. The integrated polarization levels of 70 galaxies reported by \cite{Jones_2012} are mostly below 1\% with several above 2\%.  Hence, we choose the fraction of polarized light emitted by an edge-on disk galaxy to be 2\%.

\subsection{Corrections to the toy model}
\label{sec:additional_factors}
Here we fold in additional factors that were left out in our initial assumptions. We stated earlier the assumption that our sample consists solely of edge-on spiral galaxies. However, if we consider randomly-oriented disks, only a subset would be completely edge-on. We can parameterize the inclination $i$ of  the disk galaxies with the convention that $i=0\degree$ corresponds to face-on galaxies while $i=90\degree$ corresponds to edge-on disk galaxies. The probability of observing a galaxy within $\mathrm{d}i$ of an inclination $i$ is 
\begin{equation}
P(i)\,\mathrm{d}i = \sin i\,\mathrm{d}i
\end{equation}
for $i$ between $0\degree$ and $90\degree$. This is already normalized, and hence we can calculate the expectation value of the inclination $\langle i \rangle$, which gives 1 rad. For face-on disk galaxies, the light we observe on a whole is unpolarized. If we use the $\sin^2 i$ law for the dependence of polarization effect on inclination angle in \cite{2000MNRAS.319..497S} and \cite{Jones_2012}, we can then say that the random inclinations of the disk galaxies results in $\langle e_1 \rangle$ being multiplied by a factor of $f = 0.42$.

In addition, only about 50\% of galaxies are spiral galaxies, as inferred from the visual classifications of galaxy morphology in the Great Observatories Origins Deep Survey (GOODS) fields \citep{2004ApJ...600L..93G, 2005ApJ...625..621B} into three broad morphology classes (Ellipticals, Spirals, Peculiar/Irregulars). The 2978 galaxies used in the morphology classification by \cite{2005ApJ...625..621B} are selected from the observations made by HST and the Advanced Camera for Surveys (ACS) with a magnitude limit in the $z_{850}$ band $z_{\rm AB} = 22.5$. 

We take the spiral ratio of 50\% as a rough estimate since the morphological composition is subject to the cosmic variance due to the small size of the GOODS field (0.1 deg$^2$). In addition, WFIRST will cover different wavelength bands, magnitudes and redshifts compared with ACS. However, this number will suffice for an order of magnitude estimation.

Hence, we multiply all results by a correction factor of $f= 0.21$.

\section{Polarization dependent Optical response and PSF}
\label{sect:pol_dependence}

In this section, we describe the mechanisms that induce the dependence of final image on polarizations.

\subsection{Polarization Repsonce Difference}

We start by introducing the polarization response difference $\Delta_p$. The polarization response difference $\Delta_p$ reflects the difference in the detected flux on pixels due to the polarization-dependent response of the optical system and the detection efficiency along different polarization axes. 
The choice of coordinates and the axes is arbitrary. Without loss of generality, we define the $x$-axis as the axis along which the flux of polarized light is most attenuated.
The polarization angle $\theta$ is defined against this axis.

The polarization response difference $\Delta_ p$ due to the polarization-dependent throughput has two major sources: the anti-reflective coating of the detector and the fold mirrors. Since the actual coating model of the detector is not publicly available, we estimate the plausible levels of polarization-dependent transmission of it by considering the standard single layer anti-reflective coating. We build the coating model by setting the refractive index of the substrate to be 3.2 and tuning the refractive index of the coating to reproduce the the ratio of the peak to the valley in the quantum efficiency curve of the WFIRST Wide-Field Instrument\footnote{https://wfirst.gsfc.nasa.gov/science/WFIRST\_Reference\_Information.html}. This model predicts $\Delta_p = 0.03$ at the corner of the field. WFIRST's mirrors use a protected silver coating. A bare silver fold flat would induce fractional polarization $\Delta_p=0.0014$ ($0.9$ $\mu$m) or 0.0017 ($1.93$ $\mu$m)\footnote{This uses the index of refraction model of \citet{1972PhRvB...6.4370J} and the standard Fresnel coefficients for reflection at an interface \citep[e.g.,][]{1999prop.book.....B}, and we note that $\Delta_p = |R_s-R_p|/(R_s+R_p)$, where $R_s$ and $R_p$ are the power reflection coefficients for the two polarizations.}, but coatings can significantly modify the reflection coefficients (see, e.g., \citealt{2017JATIS...3a8002H} for an analysis in the case of the Daniel K. Inouye Solar Telescope). The coatings for WFIRST at $45^\circ$ incidence have $\Delta_p$ ranging from 0.006 at 0.5 $\mu$m, decreasing to near zero at 1.1 $\mu$m, and then rising again to 0.005 at 2.4 $\mu$m.\footnote{We thank J. Kruk (private communication) for providing this information, and Harris Corporation for providing permission to publicly release the models.} The first fold mirror (F1) is at a $45^\circ$ angle and is expected to dominate the polarization, since there are much smaller angles of incidence for the second fold and the tertiary mirror. The filter may contribute too (it is at normal incidence at the center of the field, but the angle of incidence increases toward the edge), and so we plan to include it in a more complete model in the future. Our choice of $\Delta_p = 0.05$ in Table~\ref{tab:parameters} is the sum of the effects of the anti-reflection coating ($0.03$) and these three off-axis mirrors ($3\times 0.006$, with the recognition that the true effect is smaller at most wavelengths) in the optical design of the WFIRST Wide-Field Instrument.

We note that linear retardance at the fold flat will result in some conversion of Stokes $U\rightarrow V$; this phenomenon is an issue for instruments that aim to measure circular polarization. It is less of a problem here since the mechanism of polarization-dependent sensitivity (off-axis anti-reflection coating) depends on linear, not circular, polarization.

\subsection{Polarization-Induced Astigmatism}
\label{sec:astig}
When light is reflected by an optical device inside a telescope, such as the primary mirror, the reflected ray gains an extra reflectivity coefficient relative to the ray of incidence: 
$$ r = \mid r \mid e ^{\mathrm{i} \phi} .$$
The real part of the coefficient describes the relative amplitude of the reflected light, and the imaginary part describes the phase change. Both the real part and the imaginary part depend on the mirror coatings, wavelength, angle of incidence, and the polarization. The phase change by the reflection is different for s- and p-polarizations. This polarization-dependent phase change induces the polarization variation as well as wavefront aberrations to the system. We define the phase shift between the two polarizations as retardance, 
$$ \delta = \delta_s - \delta_p. $$
We can estimate the magnitude of the polarization astigmatism by adding the retardance of each reflection fold. 

The WFIRST technology report on polarization\footnote{https://wfirst.gsfc.nasa.gov/science/sdt\_public/wps/references/\\
PIAACMCstatus20161116.pdf}, which is reviewed by the Technology Assessment Committee, provides the optical modeling and linear retardance at several wavelengths. Astigmatisms has two spatial modes, vertical astigmatism (Zernike polynomial $Z^2_2$) and oblique astigmatism ($Z_2^{-2}$), which follow $\rho^2\cos (2\phi)$ and $\rho^2\sin (2\phi)$ patterns on the polar plane respectively with polar coordinate $\rho$ and angular coordinate $\phi$. At 950nm, the linear retardance is 0.005 waves (rms) for vertical astigmatism and 0.006 waves for oblique astigmatism. For our toy model, we choose 0.005 waves as a conservative estimate for both vertical and oblique astigmatisms due to the linear retardance of the fast primary mirror in the Y band of WFIRST. The WFIRST Wide-Field Instrument contains 18 Sensor Chip Assemblies (SCA). The PSF varies as a function of position in the focal plane, including across each SCA. We use the central PSF for each SCA. 
In addition to spatial variation, the PSF is also dependent on polarization. The difference in linear retardance of 0.005 waves between the two polarization directions introduces extra aberrations on top of the default PSF.

\section{Selection Bias}
\label{sect:selection_bias}

In this section, we describe how we use the ingredients in our toy model to estimate how polarization-related selection biases can affect weak lensing. 

\subsection{Defining a Galaxy Population}
\label{sect:galaxy_population}
\begin{figure}
	\centering
	\includegraphics[width=\columnwidth]{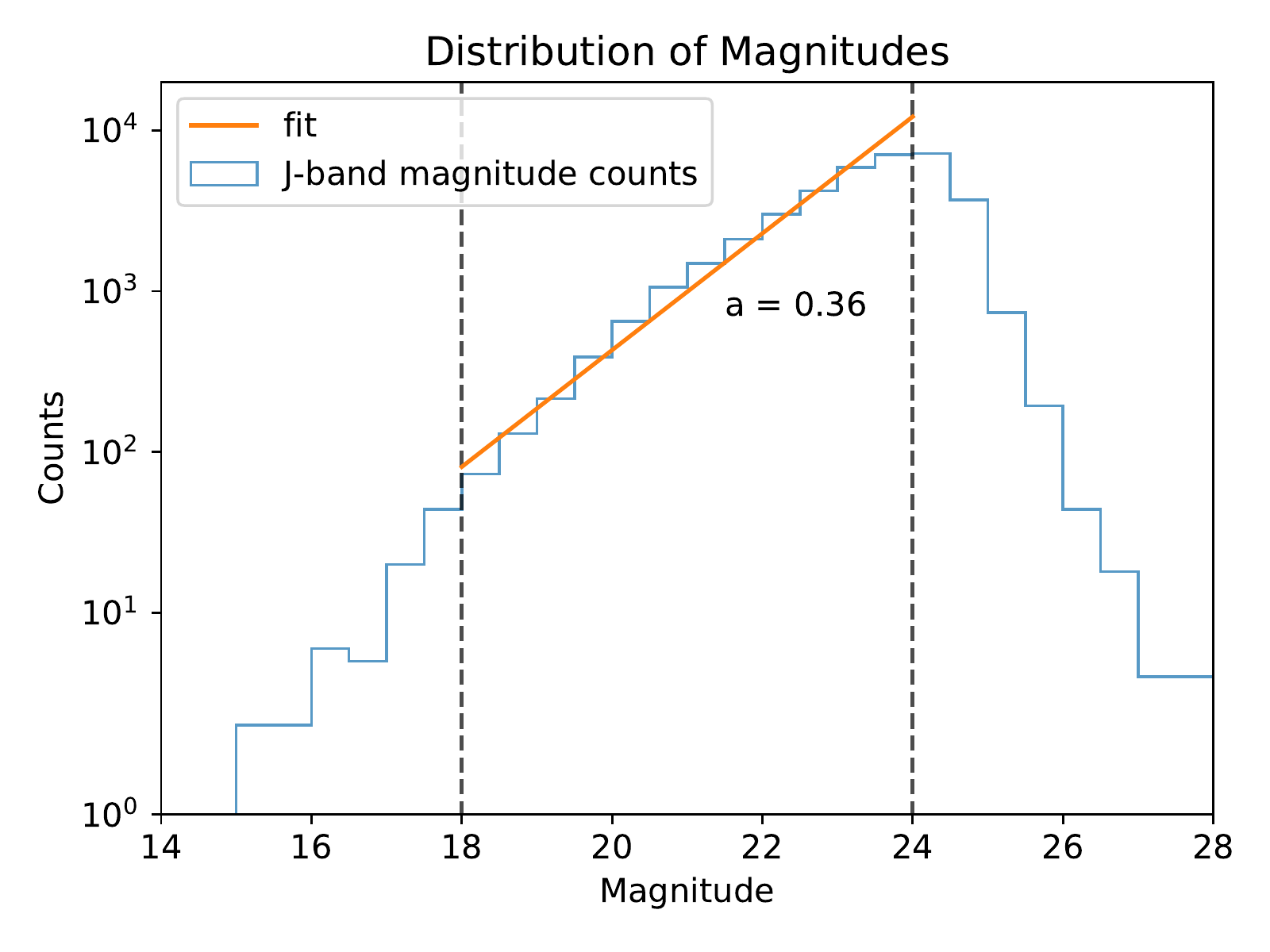}
	\caption{\label{fig:wfirst_mag_fit} This figure shows the histogram of WFIRST magnitudes $P(m)$ in J band (blue) and fitting curve (orange) of the simulated magnitude distribution to the exponential model $N \propto 10^{am}$ with $a=0.36$. The fits are valid within the magnitude range $18<m<24$. The vertical normalization of the number counts is arbitrary. The WFIRST photometry magnitudes for the four bands are simulated based on CANDELS catalogues. The dashed lines show magnitude cuts for the fits. 
	}
\end{figure}

We begin by assuming that the galaxy angles are intrinsically uniformly distributed, and that galaxy magnitudes follow a distribution $P(m)$.
To define $P(m)$, we use a simulated WFIRST photometry catalog based on CANDELS\footnote{\url{https://github.com/WFIRST-HLS-Cosmology/Docs/wiki/Home-Wiki}}. Fig.~\ref{fig:wfirst_mag_fit} shows the distribution of magnitudes in the J band. 
We then fit a rising exponential $N \propto 10^{am}$ using magnitude cuts at 18 and 24 for each band to avoid the fall-off of counts near the depth limit. 

The fitting curve is also shown in Fig.~\ref{fig:wfirst_mag_fit}. The normalization of the number counts is arbitrary. It is the slope $a$ that affects the selection bias, since it determines what fraction of the sample is sufficiently close to the boundary that it can be affected by polarization-dependent selection effects. For the magnitude distribution in the J band, the fitting within $18<m<24$ yields $a=0.36$. 

Additional corrections to the simplified toy model, for example to account for the distribution of galaxy inclination angles and the fraction of spiral galaxies, were introduced in Sect.~\ref{sec:additional_factors}.

\subsection{Effect of Polarization on Magnitude}

Following the definitions of $p$ and $\Delta_p$ in Sect. \ref{sect:pol_dependence}, we express the modification of the magnitude $m$ of a galaxy due to polarization effect via the following equation: 
\begin{equation}
m' = m - 2.5\log_{10}[(1-p) + p\sqrt{1+(b^2-1)\cos^2\theta}],
\end{equation}
where the bias factor $b$ is defined as $b = ({1 - \Delta_p})/({1 + \Delta_p})$.

Since magnitude is logarithmically related to flux, the bias factor results in a linear shift in magnitude. Hence, the term inside the logarithm represents the factor by which the flux is reduced, and has a maximal value of 1. Since $p$ represents the fraction of light that is polarized, $(1-p)$ corresponds to the unpolarized light. The square root term corresponds to the reduction in flux after the component of the flux of light polarized along the $x$-axis is multiplied by a factor of $b$, which ranges from 0 to 1. The logarithm term is negative, and hence $m'$ is greater than $m$, i.e.\ the object appears fainter due to the polarization-dependent attenuation of light.

\subsection{Magnitude Cut}
We assume that galaxies are selected by imposing a magnitude cut $m_\text{cut}$ on the galaxies such that galaxies fainter than the cut are filtered away. While this is not correct in detail \citep{2018MNRAS.481.1149Z, 2018PASJ...70S..25M}, 
it is sufficient for an order-of-magnitude estimate. The consequence of modifying galaxy magnitudes $m$ to the polarization angle-dependent $m'$ is that galaxies with magnitudes near this cut could be moved to the other side of the cut. Since $m'$ is a function of angle, the resultant population of galaxies that remain after the cut will no longer have uniformly distributed angles, violating a basic assumption in weak lensing and inducing a selection bias since an unlensed population would not have some nonzero mean ensemble shear estimate \citep{2004MNRAS.353..529H, 2005MNRAS.361.1287M}. 

We first consider the case where there is no selection bias, so $b = 1$. If we consider galaxies with a fixed polarization angle $\theta$, their magnitudes $m$ should follow the distribution $P(m)$. Hence, when we apply the magnitude cuts to the galaxy population, we can write the probability of a remaining galaxy having the polarization angle $\theta$ as:
\begin{equation}
P(\theta) \propto P(m_{\rm low}<m<m_{\rm high}) = \int_{m_{\rm low}}^{m_{\rm high}}P(m)\, \mathrm{d}m.
\end{equation}
The polarization-dependent response makes $P(\theta)$ more complicated because $m'$ is a function of $\theta$:
\begin{equation}
\label{eq:p_theta}
P(\theta) \propto P(m_{\rm low}<m'<m_{\rm high}) = \int_{m_{\rm low}+2.5\log_{10}(c)}^{m_{\rm high}+2.5\log_{10}(c)}P(m)\, \mathrm{d}m,
\end{equation}
where $c=(1-p) + p\sqrt{1+(b^2-1)\cos^2\theta}$.

\subsection{Selection bias estimation process}
We can write the expectation value of the galaxy orientation after making the cut as:
\begin{equation}
\langle \theta \rangle = \frac{\int_{0}^{\pi}\theta P(\theta)\, \mathrm{d}\theta}{\int_0^\pi P(\theta)d\theta}.
\end{equation}
Note that $\pi < \theta < 2\pi$ is degenerate with $0 < \theta < \pi$ for the polarization angle. Substituting Eq.~\eqref{eq:p_theta} gives us:
\begin{equation}
\langle\theta\rangle = \frac{1}{A} \int_{0}^{\pi}\theta
\int_{m_{\rm low}+2.5\log_{10}(c)}^{m_\text{high}+2.5\log_{10}(c)}P(m)\,\mathrm{d}m
\, \mathrm{d}\theta,
\end{equation}
where $A$ is the normalization factor,
\begin{equation}
    A = \int_{0}^{\pi}\int_{m_{\rm low}+2.5\log_{10}(c)}^{m_{\rm high}+2.5\log_{10}(c)}P(m)\,\mathrm{d}m\,\mathrm{d}\theta.
\end{equation}

As a sanity check, for the case when $c = 1$, i.e.\ the case with no polarization-dependent selection effect, the inner integral evaluates to a constant $k$ and the normalization factor is $k\pi$. Hence, we obtain $\langle\theta\rangle = \frac{\pi}{2}$ as expected for a uniform distribution of angles from 0 to $\pi$.

To quantify the effect of the selection bias on weak lensing measurements, we relate it to the two-component ellipticity $\mathbf{e} = (e_1,e_2)$, which is an observable quantity used to construct ensemble shear estimates. Making an arbitrary choice of axis with no loss of generality, we present results for the first component $\langle e_1 \rangle$.

\begin{equation}
\label{eq:e1}
\begin{split}
\langle e_1 \rangle &= \langle |\mathbf{e}|\cos(2\theta) \rangle\\
&= \frac{|\mathbf{e}|}{A} \int_{0}^{\pi}\cos(2\theta)
\int_{m_{\rm low}+2.5\log_{10}(c)}^{ m_{\rm high}+2.5\log_{10}(c)}P(m)\,\mathrm{d}m\,\mathrm{d}\theta.
\end{split}
\end{equation}

\subsection{Parameters in the simplified toy model}

We obtain an upper bound on the selection bias by choosing pessimistic conditions. The parameters that need to be specified and our values used are listed in Table~\ref{tab:parameters}.

\begin{table}
	\begin{center}
		\begin{tabular}{|c|c|c|c|c|c|}
			\hline
			Parameter & $p$ & $\Delta_p$ & $|\mathbf{e}|$ & $a$\\
			\hline
			Value & $0.02$ & $0.05$ & $0.92$ & $0.36$\\[2pt]
			\hline
		\end{tabular}
	\end{center}
	\caption{\label{tab:parameters}Choice of variables for the toy model. The variables $p$, $\Delta p$, $|\mathbf{e}|$, and $a$ are the polarization fraction, the difference of response of polarization orientations, the absolute value of ellipticity and the slope of the fitted magnitude distribution. The polarization fraction is relevant to all polarization effects, including the selection bias and the PSF errors. The other variables, $\Delta_p$, $|\mathbf{e}|$ and $a$, are specific to the selection bias. 
	}
\end{table}

In our analysis, we choose the polarization polarization to be $2\%$ and estimate $\Delta_p$ to be 5\% along the attenuated polarization axis. We choose the absolute value of ellipticity $|\mathbf{e}|$ of edge-on disk galaxies to be 0.92 based on the edge-on disk galaxy samples from the SDSS dataset \citep{2006A&A...445..765K}. The ellipticity is defined based on the second moments of galaxy image light profile throughout the paper \eqref{eq:ellipticity}. The magnitude of ellipticity is therefore $|\mathbf{e}|= (1-b^2/a^2)/(1+b^2/a^2)$ with major and minor axes $a$ and $b$ instead of $|\mathbf{e}|=1-b/a$ used in \cite{2006A&A...445..765K}. The quoted value above of the ellipticity of edge-on disk galaxies has taken into account the change of definition. As mentioned in Sect.~\ref{sect:galaxy_population}, $P(m)$ is obtained from the WFIRST simulated photometry catalog based on CANDELS. To do the experiment, we select galaxies with signal-to-noise ratio within $18 <{\rm S/N}< 200$. This leads to the magnitude cuts $(m_{\rm low}, m_{\rm high}) =(21.6, 24.2)$ in the J-band \citep{2015arXiv150303757S}, where we obtained a fit to the exponential magnitude distribution $N \propto 10^{am}$ with $a = 0.36$.

\section{PSF Errors}
\label{sect:aberration}
In this section, we describe how we use the polarization-induced aberrations introduced in Sect. \ref{sec:astig} to simulate the PSF errors for WFIRST.

\label{sect:psf}

\subsection{Image Simulation with Galsim}
In this work, we use \textsc{GalSim}\footnote{\url{https://github.com/GalSim-developers/GalSim}} ~\citep{2015A&C....10..121R} version~1.6.1 and the WFIRST module \citep{2016PASP..128i5001K} of \textsc{GalSim} to simulate the polarization-dependent PSF effect for WFIRST. 
\textsc{GalSim} is a package for simulating images of stars and galaxies. It can simulate galaxies from different galaxy models and also generate optical PSFs from parametric models. In particular, \textsc{GalSim} has a module especially designed for the image simulations for WFIRST. With \textsc{GalSim}, we simulate the PSFs for WFIRST observations for different bandpasses and Sensor Chip Assemblies (SCAs). Besides the default WFIRST aberrations, additional aberration (astigmatism) can be applied to test the polarization effect on PSFs.

\subsection{PSF shape measurement}
In order to estimate the impact of polarization on PSF errors, we measure the sizes and shapes of WFIRST PSFs with both default and polarization-dependent aberrations by considering the image of a point star. The moments of the star image are measured using the \texttt{hsm.FindAdaptiveMom} routine in \textsc{GalSim}, which calculates the adaptive moments introduced by \cite{2002AJ....123..583B} and \cite{ 2003MNRAS.343..459H}. This routine fits the 2-D image $I(\bm{x})$ with an elliptical Gaussian by iteratively minimizing the squared error 
\begin{equation}
    SE = \int \left \| I(\bm{x}) - Ae^{-\frac{1}{2}(\bm{x-\mu})^\text{T}\mathbf{M}^{-1}(\bm{x-\mu}) }\right \|^2 \mathrm{d}^2\bm{x}.
\end{equation}
The best-fit $\bm{\mu}$ and $\mathbf{M}$ are the weighted coordinates of the centroid and  moments
\begin{align}
    \bm{\mu} &= \frac{\int \bm{x} I(\bm{x}) \omega(\bm{x})\,\mathrm{d}^2 \bm{x} }{\int I(\bm{x}) \omega(\bm{x})\,\mathrm{d}^2 \bm{x}} ~~{\rm and} \nonumber \\
    \mathbf{M}_{ij} &= \frac{2 \int (\bm{x-\mu})_i(\bm{x-\mu})_j I(\bm{x}) \omega(\bm{x})\,\mathrm{d}^2 \bm{x}}{\int I(\bm{x}) \omega(\bm{x})\,\mathrm{d}^2 \bm{x}} ,
\end{align}
with the elliptic Gaussian weight function $\omega(\bm{x})$ that minimizes the squared error. 

The ellipticity components and the adaptive size $\sigma$ of the PSF are then defined in terms of the second moments: 
\begin{align}
    e_1 &= \frac{\mathbf{M}_{xx}-\mathbf{M}_{yy}}{\mathbf{M}_{xx}+\mathbf{M}_{yy}}, \nonumber \\
    e_2 &= \frac{2\mathbf{M}_{xy}}{\mathbf{M}_{xx}+\mathbf{M}_{yy}}, ~~{\rm and}\nonumber \\
    \sigma &= \sqrt[4]{\det \mathbf{M} }.
    \label{eq:ellipticity}
\end{align}
Please note that the ensemble average of this definition of ellipticity based on the second moments of the light profile, when applied to galaxies, is not an unbiased estimator of the weak lensing shear. For nearly round objects, $\langle e \rangle \approx 2 \langle \gamma \rangle$.  The requirements on our knowledge of PSF model ellipticity include the proper conversion factors to account for how systematic uncertainty in PSF model ellipticity (for this ellipticity definition) affects estimates of ensemble weak lensing shears for galaxies.

In the image simulation, we oversample the image below the WFIRST pixel scale (0.11 arcsec/pix) and down to 0.01 arcsec/pix to obtain convergent adaptive moments and adaptive sizes. The detector non-idealities, including the reciprocity failure, non-linearity and interpixel capacitance etc., are not included. We only consider the extra aberrations of PSFs due to polarization.


\section{Results}
\label{sect:results}
In this section, we present the estimated polarization-induced systematic biases using the toy model described in Sect.~\ref{sect:selection_bias} and Sect.~\ref{sect:aberration}. We show the results for both the selection bias and the PSF errors.  
\subsection{Selection bias}

Assuming an exponential magnitude distribution, we can calculate the expected mean selection bias in the shear.  
Given our fit for the exponential magnitude distribution in Sect.~\ref{sect:galaxy_population}, we can substitute into Eq.~\eqref{eq:e1} and simplify the expression. We first evaluate the inner integral,
\begin{equation}
\label{eq:P_theta}
\begin{split}
P(\theta) &\propto \int_{m_{\rm low}+2.5\log_{10}(c)}^{ m_{\rm high}+2.5\log_{10}(c)}P(m)\,\mathrm{d}m \\
&\propto [(1-p) + p\sqrt{1+(b^2-1)\cos^2\theta}]^{2.5a}.
\end{split}
\end{equation}
Going back to Eq.~\eqref{eq:e1},
\begin{equation}
\label{eq:e1_final}
\begin{split}
\langle e_1 \rangle &= \langle |\mathbf{e}|\cos(2\theta) \rangle\\
&= \frac{|\mathbf{e}|
\int_{0}^{\pi}\!
(\cos2\theta)[(1-p) + p\sqrt{1+(b^2-1)\cos^2\theta}]^{2.5a}
\mathrm{d}\theta}{\int_0^\pi [(1-p) + p\sqrt{1+(b^2-1)\cos^2\theta}]^{2.5a} d\theta}.
\end{split}
\end{equation}
This gives $\langle e_1 \rangle = 0$ when the bias parameter is 1,
as expected. For a given set of variables $p$ and $\Delta_p$, we compute this integral numerically to determine the selection bias due to polarized light profiles of edge-on disk galaxies. We note that the prefactor involving the magnitude cuts in Eq.~(\ref{eq:P_theta}) gets cancelled in Eq.~(\ref{eq:e1_final}), indicating that the expected shear component of the toy model does not depend on the magnitude cuts. In addition, the bright end limit of the magnitude cut contributes less than 10\% to the total selection effect. Setting the lower limit of the magnitude cut to negative infinity is not expected to make significant difference for the order of magnitude test.

\begin{figure}
	\centering
	\includegraphics[width=\columnwidth]{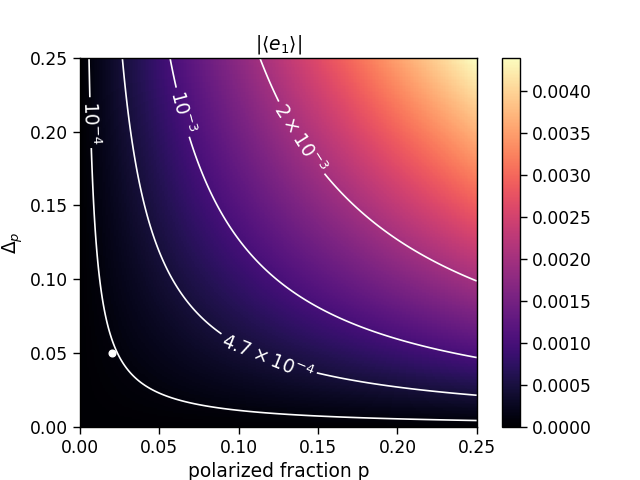}
	\caption{\label{fig:selection_bias}The expected ensemble mean shear $\langle e_1 \rangle$ component due to the selection bias for various choices of variables $p$ and $\Delta_p$. In the ideal case without a polarization-dependent optical response, $\langle e_1 \rangle$ should be zero. At our choice of parameter in Table~\ref{tab:parameters} (the white dot at the bottom left corner), the bias in $\langle e_1 \rangle$ is $\sim 8 \times 10^{-5}$. 
	}
\end{figure}

Using Eq.~\eqref{eq:e1_final} and $|\mathbf{e}| = 0.92$, we vary $p$ and $\Delta_p$ and numerically evaluate the integral, giving us the results in Fig.~\ref{fig:selection_bias}.  The additional correction factors, including the galactic inclination angle and the fraction of spiral galaxies, discussed in Sect.~\ref{sec:additional_factors}, are taken into account in this figure. For the values listed in Table~\ref{tab:parameters} of $p$ and $\Delta_p$, we obtain the estimate of $\langle e_1 \rangle = 8 \times 10^{-5}$. This is represented by the white point in Fig.~\ref{fig:selection_bias}. 

\subsection{PSF size and shape errors}

\begin{figure}
\includegraphics[width=\columnwidth]{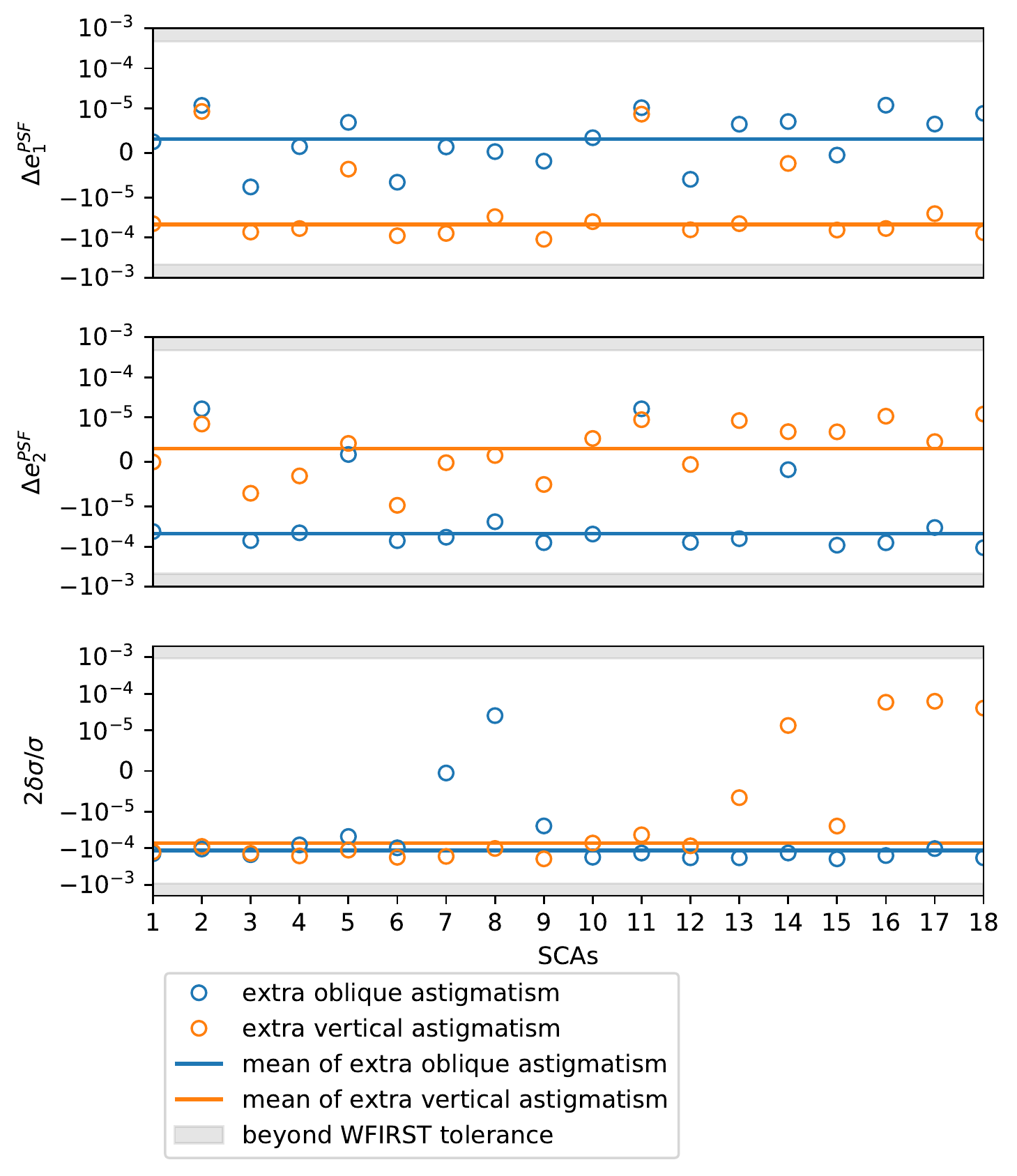}
\centering
\caption{\label{fig:psf_errors} Changes in PSF ellipticities $e_1^{\rm PSF}$ and $e_2^{\rm PSF}$ and the fractional change in the adaptive size $2(\delta \sigma)/\sigma$ due to the extra 0.005 oblique and vertical astigmatism on top of WFIRST PSF models in the Y band. The dots show the changes in each SCA, and the average offset across 18 SCAs is labelled by the solid horizontal lines. The 2\% linear polarization and the additional factors in Sect.~\ref{sec:additional_factors} are taken into account. The shaded regions are beyond the WFIRST tolerance on the ellipticity components (first two rows) and the tolerance on the relative error of trace of the second moments of the  PSFs (the third row). The polarization-induced PSF errors could account for around 20\% of the error budget. 
}
\end{figure}

To estimate the effect of light polarizations on PSF, we run the image simulation we describe in Sect. \ref{sect:psf}. In Fig.~\ref{fig:psf_errors} we show the changes in $e_1^{\rm PSF}$, $e_2^{\rm PSF}$, and the adaptive size $\sigma$ due to the extra polarization-induced aberration measured across the 18 SCAs. 
 
Table~\ref{tab:tolerance} shows the PSF errors in the Y band across the SCAs with all the factors above taken into consideration. The effect of the extra aberrations depends on the PSF of the SCAs, which is related to the positions of SCAs in the field of view. \footnote{Please see the WFIRST observatory reference information at \url{https://wfirst.gsfc.nasa.gov/science/WFIRST\_Reference\_Information.html} for the arrangement of SCAs and the field of view layout. }For the WFIRST HLS weak lensing survey, the tolerance on the relative error of trace of the second moments of the PSFs ($\delta {\rm tr}(M) / {\rm tr}(M) = 2 (\delta \sigma) / \sigma$) is $9.3 \times 10^{-4}$ and the required knowledge of PSF ellipticity is $4.7 \times 10^{-4}$ per component \citep{2015arXiv150303757S, 2016PASP..128i5001K}.
Even though the estimated PSF errors, including the error of the trace and the error of PSF ellipticity components, are within the WFIRST tolerance, the polarization-induced PSF errors alone could account for around 20\% of the error budget, indicating that more investigations into polarization-related PSFs beyond this simple toy model are required.

\begin{table*}
\centering
\begin{tabular}{llccccc}
\hline
                                     & \multicolumn{1}{c}{}               & \multicolumn{3}{c}{PSF error ($\times 10^{-4}$)} & \multirow{2}{*}{Selection bias ($\times 10^{-4}$)} & \multirow{2}{*}{WFIRST tolerance ($\times 10^{-4}$)} \\
                                     & \multicolumn{1}{c}{}               & worst          & mean           & best           &                                                    &                                                      \\ \hline
\multicolumn{1}{r}{Size Bias}        & $\delta {\rm tr}(M) / {\rm tr}(M)$ & 2.0            & -0.75          & -0.067         & -                                                  & 9.3                                                  \\
\multicolumn{1}{r}{Ellipticity Bias} & $\delta e_1$                       & -1.1           & -0.48          & 0.024          & 0.8                                                & 4.7                                                  \\ \hline
\end{tabular}
\caption{\label{tab:tolerance} This table summarizes the estimates of systematic errors due to polarization effects in our toy model. The columns show the (worst, mean, best) PSF errors due to extra polarization-induced vertical astigmatism among SCAs in Y band, the selection bias, compared to the WFIRST tolerance in the last column. 
}
\end{table*}

\section{Discussion}

In this work, we constructed a toy model to estimate the magnitude of polarization-driven systematic errors in the context of optical/NIR weak lensing surveys for the first time in the literature. We propose a simple toy model to make order of magnitude estimates for both selection bias and PSF errors due to the polarization effects. The selection bias due to the polarization-dependent optical efficiency can lead to a spurious mean shear estimate of $|\langle e_1 \rangle| \approx 8\times 10^{-5}$. For the PSF model error, our image simulations with \textsc{GalSim} in the WFIRST Y band give PSF model size and shape biases of $\delta {\rm tr}(M) / {\rm tr}(M) = 2.0 \times 10^{-4}$ and $\delta e_1^{\rm PSF}  = 1.1 \times 10^{-4}$, compared to the WFIRST requirements on the entire PSF $\delta {\rm tr}(M) / {\rm tr}(M) = 9.3 \times 10^{-4}$ and $\delta e_{1/2}^{\rm PSF}  = 4.7 \times 10^{-4}$.

There are several caveats to our toy model. The purpose of this work is to give a rough estimate of the polarization effects on weak lensing measurements, including the selection bias and PSF errors.  
Hence we adopt a simplified galaxy selection by magnitude cuts only, and use a power-law model for the galaxy number counts. We also make assumptions about the light polarization of galaxies, the dependence on galaxy inclination angle,  
the orientation of polarizations and the dependence of polarization fraction on rest frame wavelength. Some assumptions help to simplify the estimation, and some are unfortunately necessary due to the limited number of optical/NIR observations of polarizations available and/or due to what information about the WFIRST instrumentation vendors have released.

For WFIRST, the key priority will be to obtain enough measurements on both the telescope/instrument and the source galaxies to remove biases associated with polarization. The polarized detector response at non-normal incidence can be measured in the laboratory with a representative detector, and we are in discussions with the WFIRST Project Office about the best way to implement this measurement. We are also having discussions about what relevant information on the optics can be released. 
Given how faint the galaxies are, a promising way to obtain the required information on the galaxies is from WFIRST itself. In the WFIRST supernova deep fields, the same set of galaxies in a 5 deg$^2$ area are observed at $\sim 144$ epochs over 2 years \citep{2015arXiv150303757S}. In the \citet{2015arXiv150303757S} deep field strategy, the aggregate S/N for a typical lensing source (AB=24.5) is 250, and at 3 galaxies per arcmin$^2$ per redshift bin, there would be 54,000 galaxies per bin in a supernova deep field.\footnote{This calculation uses the WFIRST Exposure Time Calculator \citep{2012arXiv1204.5151H}, following \citet{2018ApJ...867...23H}, but updating with the current WFIRST Phase B throughput table, although this makes only a small difference.} The supernova fields are located near the Ecliptic poles. When observing the supernova fields, the telescope roll angle is adjusted to point its solar array toward the Sun for power. The roll angle would therefore go through a cycle every year. If there is a polarization-dependent throughput and the galaxies are polarized then we expect to see their magnitudes vary sinusoidally with a period of 6 months and a phase that corresponds to their position angle. The amplitude of this variation\footnote{Defined as half of the peak-to-valley difference; the root-mean-square is a factor of $1/\sqrt2$ smaller.} is $\approx \Delta_p\times p$. WFIRST will not measure this for any one galaxy, but if $f  \Delta_p  p$ is really $\sim 0.00021$ (where $f=0.21$ is the suppression factor in \S\ref{sec:additional_factors}), then the idealized aggregate S/N ratio on the measurement of the polarization effect is
\begin{equation}
\frac{1}{\sqrt2}\times 0.00021\times 250\times \sqrt{5.4\times 10^4} = 8.6.
\end{equation}
An advantage of this is that it directly measures the combination $\Delta_p \times p$ that is of direct interest, even without depending in the detailed model for instrument polarization $\Delta_p$. Future work should consider how to make this measurement practical, particularly when incorporating large-scale flat field uncertainties and dithering strategies. 

In this paper, the biases due to polarization-dependent throughput and PSF are both estimated with WFIRST-specific parameters. The polarization biases for surveys with different optical design, targeted wavelength and point spread function would be different. In addition, most weak lensing surveys to date are ground-based and thus subject to atmospheric turbulence. It is nontrivial to infer the impact of polarization effect for other surveys without tailored analysis. But for WFIRST, both of the polarization-induced errors are comparable to the WFIRST tolerance. We thus recommend more complete studies on polarization-induced systematic uncertainties of weak lensing measurements. 

\label{sect:conclusions}
\section*{Acknowledgments}

We would like to thank the WFIRST Project Office for useful information and input. 
Chien-Hao Lin, Rachel Mandelbaum, and Christopher Hirata carried out this work as part of the WFIRST Science Investigation Team ``Cosmology with the High Latitude Survey''  supported by NASA grant 15-WFIRST15-0008. 
The author Brent Tan appreciates support from NASA grant NNX17AK58G.
During the preparation of this work, CMH has also been supported by the Simons Foundation award 60052667 and US Department of Energy award DE-SC0019083.

\bibliography{pol}

\begin{thebibliography}{}
\makeatletter
\relax
\def\mn@urlcharsother{\let\do\@makeother \do\$\do\&\do\#\do\^\do\_\do\%\do\~}
\def\mn@doi{\begingroup\mn@urlcharsother \@ifnextchar [ {\mn@doi@}
  {\mn@doi@[]}}
\def\mn@doi@[#1]#2{\def\@tempa{#1}\ifx\@tempa\@empty \href
  {http://dx.doi.org/#2} {doi:#2}\else \href {http://dx.doi.org/#2} {#1}\fi
  \endgroup}
\def\mn@eprint#1#2{\mn@eprint@#1:#2::\@nil}
\def\mn@eprint@arXiv#1{\href {http://arxiv.org/abs/#1} {{\tt arXiv:#1}}}
\def\mn@eprint@dblp#1{\href {http://dblp.uni-trier.de/rec/bibtex/#1.xml}
  {dblp:#1}}
\def\mn@eprint@#1:#2:#3:#4\@nil{\def\@tempa {#1}\def\@tempb {#2}\def\@tempc
  {#3}\ifx \@tempc \@empty \let \@tempc \@tempb \let \@tempb \@tempa \fi \ifx
  \@tempb \@empty \def\@tempb {arXiv}\fi \@ifundefined
  {mn@eprint@\@tempb}{\@tempb:\@tempc}{\expandafter \expandafter \csname
  mn@eprint@\@tempb\endcsname \expandafter{\@tempc}}}

\bibitem[\protect\citeauthoryear{{Akahori} et~al.,}{{Akahori}
  et~al.}{2018}]{2018PASJ...70R...2A}
{Akahori} T.,  et~al., 2018, \mn@doi [\pasj] {10.1093/pasj/psx123}, \href
  {http://adsabs.harvard.edu/abs/2018PASJ...70R...2A} {70, R2}

\bibitem[\protect\citeauthoryear{{Andersson}, {Lazarian}  \&
  {Vaillancourt}}{{Andersson} et~al.}{2015}]{2015ARA&A..53..501A}
{Andersson} B.~G.,  {Lazarian} A.,   {Vaillancourt} J.~E.,  2015, \mn@doi
  [\araa] {10.1146/annurev-astro-082214-122414}, \href
  {https://ui.adsabs.harvard.edu/abs/2015ARA&A..53..501A} {53, 501}

\bibitem[\protect\citeauthoryear{{Bartelmann} \& {Schneider}}{{Bartelmann} \&
  {Schneider}}{2001}]{2001PhR...340..291B}
{Bartelmann} M.,  {Schneider} P.,  2001, \mn@doi [\physrep]
  {10.1016/S0370-1573(00)00082-X}, \href
  {http://adsabs.harvard.edu/abs/2001PhR...340..291B} {340, 291}

\bibitem[\protect\citeauthoryear{{Beck}}{{Beck}}{2007}]{2007EAS....23...19B}
{Beck} R.,  2007, in {Miville-Desch{\^e}nes} M.~A.,  {Boulanger} F.,  eds,  EAS
  Publications Series Vol. 23, EAS Publications Series. pp 19--36 (\mn@eprint
  {arXiv} {astro-ph/0603531}), \mn@doi{10.1051/eas:2007003}

\bibitem[\protect\citeauthoryear{{Beck}, {Shoutenkov}, {Ehle}, {Harnett},
  {Haynes}, {Shukurov}, {Sokoloff}  \& {Thierbach}}{{Beck}
  et~al.}{2002}]{2002A&A...391...83B}
{Beck} R.,  {Shoutenkov} V.,  {Ehle} M.,  {Harnett} J.~I.,  {Haynes} R.~F.,
  {Shukurov} A.,  {Sokoloff} D.~D.,   {Thierbach} M.,  2002, \mn@doi [\aap]
  {10.1051/0004-6361:20020642}, \href
  {https://ui.adsabs.harvard.edu/abs/2002A&A...391...83B} {391, 83}

\bibitem[\protect\citeauthoryear{{Bernstein} \& {Jarvis}}{{Bernstein} \&
  {Jarvis}}{2002}]{2002AJ....123..583B}
{Bernstein} G.~M.,  {Jarvis} M.,  2002, \mn@doi [\aj] {10.1086/338085}, \href
  {http://adsabs.harvard.edu/abs/2002AJ....123..583B} {123, 583}

\bibitem[\protect\citeauthoryear{{Born} \& {Wolf}}{{Born} \&
  {Wolf}}{1999}]{1999prop.book.....B}
{Born} M.,  {Wolf} E.,  1999, {Principles of Optics}

\bibitem[\protect\citeauthoryear{{Breckinridge}, {Lam}  \&
  {Chipman}}{{Breckinridge} et~al.}{2015}]{2015PASP..127..445B}
{Breckinridge} J.~B.,  {Lam} W. S.~T.,   {Chipman} R.~A.,  2015, \mn@doi
  [\pasp] {10.1086/681280}, \href
  {https://ui.adsabs.harvard.edu/abs/2015PASP..127..445B} {127, 445}

\bibitem[\protect\citeauthoryear{{Bundy}, {Ellis}  \& {Conselice}}{{Bundy}
  et~al.}{2005}]{2005ApJ...625..621B}
{Bundy} K.,  {Ellis} R.~S.,   {Conselice} C.~J.,  2005, \mn@doi [\apj]
  {10.1086/429549}, \href {http://adsabs.harvard.edu/abs/2005ApJ...625..621B}
  {625, 621}

\bibitem[\protect\citeauthoryear{{Chipman}, {Lam}  \& {Breckinridge}}{{Chipman}
  et~al.}{2015}]{RC_polarization2015}
{Chipman} R.~A.,  {Lam} W. S.~T.,   {Breckinridge} J.,  2015, Proc. SPIE

\bibitem[\protect\citeauthoryear{{Davis} \& {Greenstein}}{{Davis} \&
  {Greenstein}}{1951}]{1951ApJ...114..206D}
{Davis} Jr. L.,  {Greenstein} J.~L.,  1951, \mn@doi [\apj] {10.1086/145464},
  \href {http://adsabs.harvard.edu/abs/1951ApJ...114..206D} {114, 206}

\bibitem[\protect\citeauthoryear{{Draine}}{{Draine}}{2003}]{2003ApJ...598.1017D}
{Draine} B.~T.,  2003, \mn@doi [\apj] {10.1086/379118}, \href
  {https://ui.adsabs.harvard.edu/abs/2003ApJ...598.1017D} {598, 1017}

\bibitem[\protect\citeauthoryear{{Draper}, {Done}, {Scarrott}  \&
  {Stockdale}}{{Draper} et~al.}{1995}]{1995MNRAS.277.1430D}
{Draper} P.~W.,  {Done} C.,  {Scarrott} S.~M.,   {Stockdale} D.~P.,  1995,
  \mn@doi [\mnras] {10.1093/mnras/277.4.1430}, \href
  {http://adsabs.harvard.edu/abs/1995MNRAS.277.1430D} {277, 1430}

\bibitem[\protect\citeauthoryear{{Fendt}, {Beck}, {Lesch}  \&
  {Neininger}}{{Fendt} et~al.}{1996}]{1996A&A...308..713F}
{Fendt} C.,  {Beck} R.,  {Lesch} H.,   {Neininger} N.,  1996, \aap, \href
  {http://adsabs.harvard.edu/abs/1996A\%26A...308..713F} {308, 713}

\bibitem[\protect\citeauthoryear{{Fendt}, {Beck}  \& {Neininger}}{{Fendt}
  et~al.}{1998}]{1998A&A...335..123F}
{Fendt} C.,  {Beck} R.,   {Neininger} N.,  1998, \aap, \href
  {http://adsabs.harvard.edu/abs/1998A\%26A...335..123F} {335, 123}

\bibitem[\protect\citeauthoryear{{Giavalisco} et~al.,}{{Giavalisco}
  et~al.}{2004}]{2004ApJ...600L..93G}
{Giavalisco} M.,  et~al., 2004, \mn@doi [\apjl] {10.1086/379232}, \href
  {http://adsabs.harvard.edu/abs/2004ApJ...600L..93G} {600, L93}

\bibitem[\protect\citeauthoryear{{Harrington} \& {Sueoka}}{{Harrington} \&
  {Sueoka}}{2017}]{2017JATIS...3a8002H}
{Harrington} D.~M.,  {Sueoka} S.~R.,  2017, \mn@doi [Journal of Astronomical
  Telescopes, Instruments, and Systems] {10.1117/1.JATIS.3.1.018002}, \href
  {https://ui.adsabs.harvard.edu/abs/2017JATIS...3a8002H} {3, 018002}

\bibitem[\protect\citeauthoryear{{Hikage} et~al.,}{{Hikage}
  et~al.}{2019}]{2019PASJ..tmp...22H}
{Hikage} C.,  et~al., 2019, \mn@doi [Publications of the Astronomical Society
  of Japan] {10.1093/pasj/psz010}, \href
  {https://ui.adsabs.harvard.edu/abs/2019PASJ..tmp...22H} {p.~22}

\bibitem[\protect\citeauthoryear{{Hildebrandt} et~al.,}{{Hildebrandt}
  et~al.}{2018}]{2018arXiv181206076H}
{Hildebrandt} H.,  et~al., 2018, arXiv e-prints, \href
  {https://ui.adsabs.harvard.edu/\#abs/2018arXiv181206076H} {p.
  arXiv:1812.06076}

\bibitem[\protect\citeauthoryear{{Hirata} \& {Seljak}}{{Hirata} \&
  {Seljak}}{2003}]{2003MNRAS.343..459H}
{Hirata} C.,  {Seljak} U.,  2003, \mn@doi [\mnras]
  {10.1046/j.1365-8711.2003.06683.x}, \href
  {http://adsabs.harvard.edu/abs/2003MNRAS.343..459H} {343, 459}

\bibitem[\protect\citeauthoryear{{Hirata} et~al.,}{{Hirata}
  et~al.}{2004}]{2004MNRAS.353..529H}
{Hirata} C.~M.,  et~al., 2004, \mn@doi [\mnras]
  {10.1111/j.1365-2966.2004.08090.x}, \href
  {https://ui.adsabs.harvard.edu/\#abs/2004MNRAS.353..529H} {353, 529}

\bibitem[\protect\citeauthoryear{{Hirata}, {Gehrels}, {Kneib}, {Kruk},
  {Rhodes}, {Wang}  \& {Zoubian}}{{Hirata} et~al.}{2012}]{2012arXiv1204.5151H}
{Hirata} C.~M.,  {Gehrels} N.,  {Kneib} J.-P.,  {Kruk} J.,  {Rhodes} J.,
  {Wang} Y.,   {Zoubian} J.,  2012, arXiv e-prints, \href
  {https://ui.adsabs.harvard.edu/abs/2012arXiv1204.5151H} {p. arXiv:1204.5151}

\bibitem[\protect\citeauthoryear{{Hoekstra} \& {Jain}}{{Hoekstra} \&
  {Jain}}{2008}]{2008ARNPS..58...99H}
{Hoekstra} H.,  {Jain} B.,  2008, \mn@doi [Annual Review of Nuclear and
  Particle Science] {10.1146/annurev.nucl.58.110707.171151}, \href
  {http://adsabs.harvard.edu/abs/2008ARNPS..58...99H} {58, 99}

\bibitem[\protect\citeauthoryear{{Hounsell} et~al.,}{{Hounsell}
  et~al.}{2018}]{2018ApJ...867...23H}
{Hounsell} R.,  et~al., 2018, \mn@doi [\apj] {10.3847/1538-4357/aac08b}, \href
  {https://ui.adsabs.harvard.edu/abs/2018ApJ...867...23H} {867, 23}

\bibitem[\protect\citeauthoryear{{Huff}, {Eifler}, {Hirata}, {Mandelbaum},
  {Schlegel}  \& {Seljak}}{{Huff} et~al.}{2014}]{heh14}
{Huff} E.~M.,  {Eifler} T.,  {Hirata} C.~M.,  {Mandelbaum} R.,  {Schlegel} D.,
   {Seljak} U.,  2014, \mn@doi [\mnras] {10.1093/mnras/stu145}, \href
  {http://adsabs.harvard.edu/abs/2014MNRAS.440.1322H} {440, 1322}

\bibitem[\protect\citeauthoryear{{Ivezi{\'c}} et~al.,}{{Ivezi{\'c}}
  et~al.}{2008}]{2008arXiv0805.2366I}
{Ivezi{\'c}} {\v{Z}}.,  et~al., 2008, arXiv e-prints, \href
  {https://ui.adsabs.harvard.edu/\#abs/2008arXiv0805.2366I} {p.
  arXiv:0805.2366}

\bibitem[\protect\citeauthoryear{{Jee}, {Tyson}, {Hilbert}, {Schneider},
  {Schmidt}  \& {Wittman}}{{Jee} et~al.}{2016}]{2016ApJ...824...77J}
{Jee} M.~J.,  {Tyson} J.~A.,  {Hilbert} S.,  {Schneider} M.~D.,  {Schmidt} S.,
   {Wittman} D.,  2016, \mn@doi [\apj] {10.3847/0004-637X/824/2/77}, \href
  {https://ui.adsabs.harvard.edu/\#abs/2016ApJ...824...77J} {824, 77}

\bibitem[\protect\citeauthoryear{{Johnson} \& {Christy}}{{Johnson} \&
  {Christy}}{1972}]{1972PhRvB...6.4370J}
{Johnson} P.~B.,  {Christy} R.~W.,  1972, \mn@doi [\prb]
  {10.1103/PhysRevB.6.4370}, \href
  {https://ui.adsabs.harvard.edu/abs/1972PhRvB...6.4370J} {6, 4370}

\bibitem[\protect\citeauthoryear{{Jones}}{{Jones}}{2000}]{2000AJ....120.2920J}
{Jones} T.~J.,  2000, \mn@doi [\aj] {10.1086/316880}, \href
  {http://adsabs.harvard.edu/abs/2000AJ....120.2920J} {120, 2920}

\bibitem[\protect\citeauthoryear{Jones, Wang, Krisciunas  \& Freeland}{Jones
  et~al.}{2012}]{Jones_2012}
Jones A.,  Wang L.,  Krisciunas K.,   Freeland E.,  2012, \mn@doi [The
  Astrophysical Journal] {10.1088/0004-637x/748/1/17}, 748, 17

\bibitem[\protect\citeauthoryear{{Jura}}{{Jura}}{1982}]{1982ApJ...258...59J}
{Jura} M.,  1982, \mn@doi [\apj] {10.1086/160052}, \href
  {http://adsabs.harvard.edu/abs/1982ApJ...258...59J} {258, 59}

\bibitem[\protect\citeauthoryear{{Kannawadi}, {Shapiro}, {Mandelbaum},
  {Hirata}, {Kruk}  \& {Rhodes}}{{Kannawadi}
  et~al.}{2016}]{2016PASP..128i5001K}
{Kannawadi} A.,  {Shapiro} C.~A.,  {Mandelbaum} R.,  {Hirata} C.~M.,  {Kruk}
  J.~W.,   {Rhodes} J.~D.,  2016, \mn@doi [\pasp]
  {10.1088/1538-3873/128/967/095001}, \href
  {http://adsabs.harvard.edu/abs/2016PASP..128i5001K} {128, 095001}

\bibitem[\protect\citeauthoryear{{Kautsch}, {Grebel}, {Barazza}  \&
  {Gallagher}}{{Kautsch} et~al.}{2006}]{2006A&A...445..765K}
{Kautsch} S.~J.,  {Grebel} E.~K.,  {Barazza} F.~D.,   {Gallagher} J.~S. I.,
  2006, \mn@doi [\aap] {10.1051/0004-6361:20053981}, \href
  {https://ui.adsabs.harvard.edu/abs/2006A&A...445..765K} {445, 765}

\bibitem[\protect\citeauthoryear{{Kilbinger}}{{Kilbinger}}{2015}]{2015RPPh...78h6901K}
{Kilbinger} M.,  2015, \mn@doi [Reports on Progress in Physics]
  {10.1088/0034-4885/78/8/086901}, \href
  {http://adsabs.harvard.edu/abs/2015RPPh...78h6901K} {78, 086901}

\bibitem[\protect\citeauthoryear{{LSST Science Collaboration} et~al.,}{{LSST
  Science Collaboration} et~al.}{2009}]{2009arXiv0912.0201L}
{LSST Science Collaboration} et~al., 2009, preprint, \href
  {http://adsabs.harvard.edu/abs/2009arXiv0912.0201L} {} (\mn@eprint {arXiv}
  {0912.0201})

\bibitem[\protect\citeauthoryear{{Laureijs} et~al.,}{{Laureijs}
  et~al.}{2011}]{2011arXiv1110.3193L}
{Laureijs} R.,  et~al., 2011, preprint, \href
  {http://adsabs.harvard.edu/abs/2011arXiv1110.3193L} {} (\mn@eprint {arXiv}
  {1110.3193})

\bibitem[\protect\citeauthoryear{Mandelbaum}{Mandelbaum}{2018}]{doi:10.1146/annurev-astro-081817-051928}
Mandelbaum R.,  2018, \mn@doi [Annual Review of Astronomy and Astrophysics]
  {10.1146/annurev-astro-081817-051928}, 56, 393

\bibitem[\protect\citeauthoryear{{Mandelbaum} et~al.,}{{Mandelbaum}
  et~al.}{2005}]{2005MNRAS.361.1287M}
{Mandelbaum} R.,  et~al., 2005, \mn@doi [\mnras]
  {10.1111/j.1365-2966.2005.09282.x}, \href
  {https://ui.adsabs.harvard.edu/\#abs/2005MNRAS.361.1287M} {361, 1287}

\bibitem[\protect\citeauthoryear{{Mandelbaum} et~al.,}{{Mandelbaum}
  et~al.}{2018}]{2018PASJ...70S..25M}
{Mandelbaum} R.,  et~al., 2018, \mn@doi [Publications of the Astronomical
  Society of Japan] {10.1093/pasj/psx130}, \href
  {https://ui.adsabs.harvard.edu/\#abs/2018PASJ...70S..25M} {70, S25}

\bibitem[\protect\citeauthoryear{{Massey} et~al.,}{{Massey}
  et~al.}{2007}]{2007ApJS..172..239M}
{Massey} R.,  et~al., 2007, \mn@doi [The Astrophysical Journal Supplement
  Series] {10.1086/516599}, \href
  {https://ui.adsabs.harvard.edu/\#abs/2007ApJS..172..239M} {172, 239}

\bibitem[\protect\citeauthoryear{{Mathis}}{{Mathis}}{1990}]{1990ARA&A..28...37M}
{Mathis} J.~S.,  1990, \mn@doi [\araa] {10.1146/annurev.aa.28.090190.000345},
  \href {http://adsabs.harvard.edu/abs/1990ARA%26A..28...37M} {28, 37}

\bibitem[\protect\citeauthoryear{{Radhakrishnan}}{{Radhakrishnan}}{1989}]{1989QJRAS..30..181R}
{Radhakrishnan} V.,  1989, \qjras, \href
  {http://adsabs.harvard.edu/abs/1989QJRAS..30..181R} {30, 181}

\bibitem[\protect\citeauthoryear{{Rowe} et~al.,}{{Rowe}
  et~al.}{2015}]{2015A&C....10..121R}
{Rowe} B.~T.~P.,  et~al., 2015, \mn@doi [Astronomy and Computing]
  {10.1016/j.ascom.2015.02.002}, \href
  {http://adsabs.harvard.edu/abs/2015A\%26C....10..121R} {10, 121}

\bibitem[\protect\citeauthoryear{{Scarrott}}{{Scarrott}}{1996}]{1996QJRAS..37..297S}
{Scarrott} S.~M.,  1996, \qjras, \href
  {http://adsabs.harvard.edu/abs/1996QJRAS..37..297S} {37, 297}

\bibitem[\protect\citeauthoryear{{Scarrott}, {Ward-Thompson}  \&
  {Warren-Smith}}{{Scarrott} et~al.}{1987}]{1987MNRAS.224..299S}
{Scarrott} S.~M.,  {Ward-Thompson} D.,   {Warren-Smith} R.~F.,  1987, \mn@doi
  [\mnras] {10.1093/mnras/224.2.299}, \href
  {http://adsabs.harvard.edu/abs/1987MNRAS.224..299S} {224, 299}

\bibitem[\protect\citeauthoryear{{Scarrott}, {Rolph}  \& {Semple}}{{Scarrott}
  et~al.}{1990}]{1990IAUS..140..245S}
{Scarrott} S.~M.,  {Rolph} C.~D.,   {Semple} D.~P.,  1990, in {Beck} R.,
  {Kronberg} P.~P.,   {Wielebinski} R.,  eds,  IAU Symposium Vol. 140, Galactic
  and Intergalactic Magnetic Fields. pp 245--251

\bibitem[\protect\citeauthoryear{{Scarrott}, {Eaton}  \& {Axon}}{{Scarrott}
  et~al.}{1991}]{1991MNRAS.252P..12S}
{Scarrott} S.~M.,  {Eaton} N.,   {Axon} D.~J.,  1991, \mn@doi [\mnras]
  {10.1093/mnras/252.1.12P}, \href
  {http://adsabs.harvard.edu/abs/1991MNRAS.252P..12S} {252, 12P}

\bibitem[\protect\citeauthoryear{{Simmons} \& {Audit}}{{Simmons} \&
  {Audit}}{2000}]{2000MNRAS.319..497S}
{Simmons} J.~F.~L.,  {Audit} E.,  2000, \mn@doi [\mnras]
  {10.1046/j.1365-8711.2000.03863.x}, \href
  {http://adsabs.harvard.edu/abs/2000MNRAS.319..497S} {319, 497}

\bibitem[\protect\citeauthoryear{{Sofue}, {Fujimoto}  \& {Wielebinski}}{{Sofue}
  et~al.}{1986}]{1986ARA&A..24..459S}
{Sofue} Y.,  {Fujimoto} M.,   {Wielebinski} R.,  1986, \mn@doi [\araa]
  {10.1146/annurev.aa.24.090186.002331}, \href
  {http://adsabs.harvard.edu/abs/1986ARA\%26A..24..459S} {24, 459}

\bibitem[\protect\citeauthoryear{{Spergel} et~al.,}{{Spergel}
  et~al.}{2015}]{2015arXiv150303757S}
{Spergel} D.,  et~al., 2015, preprint, \href
  {http://adsabs.harvard.edu/abs/2015arXiv150303757S} {} (\mn@eprint {arXiv}
  {1503.03757})

\bibitem[\protect\citeauthoryear{{Spitzer}}{{Spitzer}}{1978}]{1978ppim.book.....S}
{Spitzer} L.,  1978, {Physical processes in the interstellar medium},
  \mn@doi{10.1002/9783527617722.
}

\bibitem[\protect\citeauthoryear{Stil, Krause, Beck  \& Taylor}{Stil
  et~al.}{2009}]{Stil_2009}
Stil J.~M.,  Krause M.,  Beck R.,   Taylor A.~R.,  2009, \mn@doi [The
  Astrophysical Journal] {10.1088/0004-637x/693/2/1392}, 693, 1392

\bibitem[\protect\citeauthoryear{{Troxel} et~al.,}{{Troxel}
  et~al.}{2018}]{2018PhRvD..98d3528T}
{Troxel} M.~A.,  et~al., 2018, \mn@doi [\prd] {10.1103/PhysRevD.98.043528},
  \href {https://ui.adsabs.harvard.edu/\#abs/2018PhRvD..98d3528T} {98, 043528}

\bibitem[\protect\citeauthoryear{{Weingartner} \& {Draine}}{{Weingartner} \&
  {Draine}}{2001}]{2001ApJ...548..296W}
{Weingartner} J.~C.,  {Draine} B.~T.,  2001, \mn@doi [\apj] {10.1086/318651},
  \href {https://ui.adsabs.harvard.edu/abs/2001ApJ...548..296W} {548, 296}

\bibitem[\protect\citeauthoryear{{Zuntz} et~al.,}{{Zuntz}
  et~al.}{2018}]{2018MNRAS.481.1149Z}
{Zuntz} J.,  et~al., 2018, \mn@doi [\mnras] {10.1093/mnras/sty2219}, \href
  {https://ui.adsabs.harvard.edu/\#abs/2018MNRAS.481.1149Z} {481, 1149}

\makeatother
\end{thebibliography}

\end{document}